# Deep-learning-enabled rapid spatially multiplexed wavefront sensing with a thin diffuser

Ming Xue[1,2,†], Shijie Wei[1,†], Shulei Zhao[1,2], Zhuojia Shi[1,2], Tengfei Wu[1,2,*] and Xiaopeng Shao[1,2]

[1]Frontier Research Center, Xi'an Institute of Optics and Precision Mechanics, Chinese Academy of Sciences, Xi'an 710119, China

[2]University of Chinese Academy of Sciences, Beijing 100049, China

[†]These authors contributed equally to this work.

[*]wutengfei@opt.ac.cn

**Abstract**

Conventional wavefront sensing typically measures aberrations along a single guide-star direction, limiting the correction of field-dependent aberrations over a wide field of view. Spatial multiplexing with a thin diffuser enables simultaneous measurements across multiple isoplanatic patches, but the computational cost of reconstruction makes it difficult to track changing aberrations. We propose a deep-learning approach that combines recurrent all-pairs field transforms for displacement estimation with Zernike coefficient regression. Experimental results show that the method accurately reconstructs multiple wavefronts from a single speckle pattern and reduces three-channel reconstruction time from 2801.6 ms with conventional digital image correlation (DIC) to 23.1 ms. The reconstructed wavefronts agreed more closely with sequential single-channel references than those obtained by multiplexed DIC, and simulations showed that Zernike regression was more robust to noise than direct gradient integration. These results suggest a path toward dynamic wide-field aberration sensing and high-speed adaptive optics.

## Introduction

Optical aberrations induced by inhomogeneous media degrade image quality in astronomy and microscopy [1]. Adaptive optics (AO) compensates for these aberrations using wavefront information obtained through either indirect or direct sensing. Indirect, or sensorless, wavefront sensing estimates aberrations through multiple image acquisitions and iterative optimization, avoiding dedicated wavefront sensors but requiring longer measurement times and potentially converging to local minima [2,3]. Direct wavefront sensing enables rapid measurements using a guide star and a wavefront sensor, such as a Shack–Hartmann sensor, at the cost of additional optical components [4,5]. However, when aberrations vary across the field of view, a wavefront measured from a single guide star is generally representative only within a limited isoplanatic patch, restricting the effective AO correction field [6].

Several approaches have been developed to extend the correction field. Multi-conjugate AO uses multiple guide stars and correction elements conjugated to different planes to compensate for aberrations at different depths, but increases optical and calibration complexity and remains limited by the number of correction planes [7,8]. Sequential pupil modulation recovers field-dependent aberrations by acquiring images at successive subaperture positions [9,10], whereas scanning guide-star methods measure wavefronts sequentially across different field directions [11]. Both approaches require multiple acquisitions, making simultaneous characterization of rapidly varying aberrations difficult. These limitations motivate wavefront sensing across multiple field directions in a single exposure. Holographic diffuser-based wavefront sensing (HD-WFS) exploits the angular memory effect for quantitative phase measurement [12].

Subsequent spatially multiplexed approaches have enabled simultaneous reconstruction of wavefronts from different isoplanatic patches using a single speckle image [13]. However, iterative digital image correlation (DIC) is computationally demanding [14], limiting the use of these methods for time-sensitive applications such as dynamic aberration sensing.

We propose a deep-learning-based framework for rapid multiplexed wavefront reconstruction from a single speckle pattern, substantially enhancing the measurement efficiency of HD-WFS. In HD-WFS, wavefront aberrations are encoded in the displacements of speckle patterns relative to their corresponding reference patterns [12,13], making dense displacement estimation a natural fit for deep-learning-based feature extraction [15–18]. Figure 1 illustrates the two stages of the proposed framework. First, channel-specific residual speckle patterns are generated by subtracting the reference patterns associated with the other channels from the multiplexed measurement. A dense optical flow network based on the recurrent all-pairs field transforms (RAFT) architecture [19], fine-tuned on HD-WFS speckle data, is then employed to estimate the displacement field for each channel. Second, a convolutional neural network is trained to directly map the estimated displacement fields and speckle intensity features to Zernike coefficients [20,21], thereby enabling wavefront reconstruction without explicit gradient integration. Experimental results demonstrate that the proposed method reduces the reconstruction time from 2801.6 to 23.1 ms, achieving an approximately 121-fold speedup over the conventional DIC method while maintaining reconstruction accuracy. Furthermore, the Zernike regression stage exhibits substantially improved reconstruction robustness under noisy conditions compared with direct numerical integration. These results establish a computationally efficient framework for reconstructing dynamically varying field-dependent aberrations and provide a foundation for further exploration of wavefront sensing under low-signal-to-noise-ratio (SNR) and faint-target conditions.

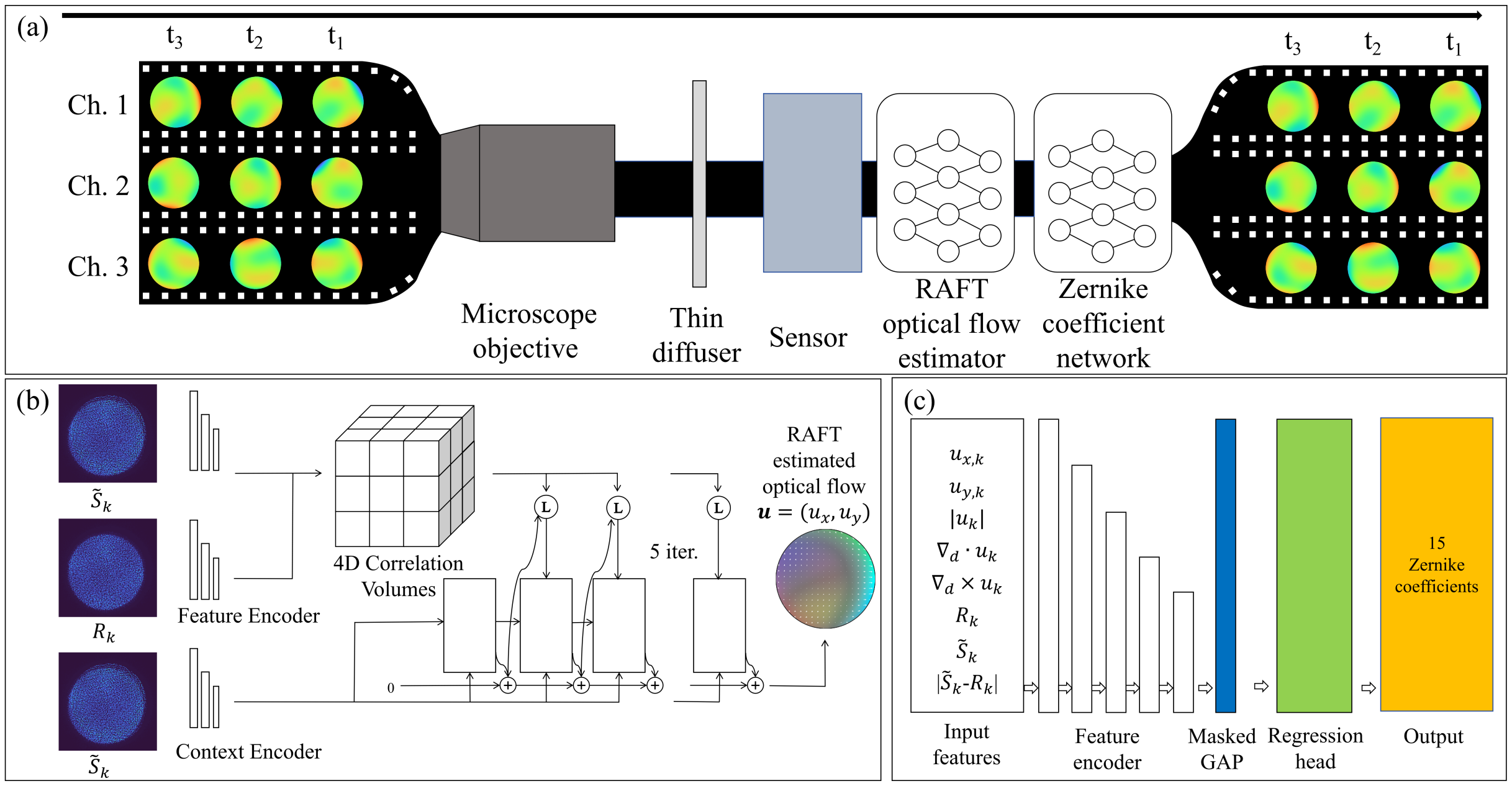

**Fig. 1.** Schematic of the proposed spatially multiplexed wavefront sensing framework. (a) Optical configuration and two-stage deep-learning reconstruction pipeline. (b) RAFT-based displacement estimation using four-dimensional (4D) reference–residual feature correlations and five recurrent updates. (c) Convolutional regression network mapping eight co-registered displacement and speckle intensity features to 15 non-piston Zernike coefficients. Masked GAP denotes pupil-masked global average pooling.

## Principle

Within the diffuser's angular memory-effect range, local wavefront tilts shift the speckle pattern while largely preserving its spatial structure [12,13]. In spatially multiplexed measurements, patterns from different field directions overlap in one camera image, complicating channel-specific displacement estimation. We recover the wavefronts from this superposition by combining channel-specific residual formation, dense displacement estimation, and Zernike coefficient regression (Fig. 1).

For channel $k$ ($k$=1,...,$N$), let $R_k(\boldsymbol{r})$ denote the unaberrated reference speckle pattern at pixel coordinates $\boldsymbol{r}$, acquired on a common camera intensity scale without frame-wise normalization. The corresponding aberrated speckle pattern is approximated by

$$S_k(\boldsymbol{r})=I_k(\boldsymbol{r})\,R_k[\boldsymbol{r}-\boldsymbol{u}_k(\boldsymbol{r})] \tag{1}$$

where $S_k(\boldsymbol{r})$ is the aberrated channel intensity, $I_k(\boldsymbol{r})$ is a slowly varying relative modulation, and $\boldsymbol{u}_k(\boldsymbol{r})$ is the reference-relative pixel displacement. Under the small-angle approximation, the transverse gradient of the optical path difference (OPD), $W_k(\boldsymbol{r})$, satisfies:

$$\nabla_{\perp} W_k(\boldsymbol{r})\simeq\frac{p}{d}\,\boldsymbol{u}_k(\boldsymbol{r}) \tag{2}$$

where $p$ is the pixel pitch, $d$ is the effective diffuser-to-sensor distance, and $\nabla_{\perp}$ is evaluated at $\boldsymbol{x}=p\boldsymbol{r}$. For $N$ mutually incoherent channels, the recorded multiplexed intensity is

$$M(\boldsymbol{r})=\sum_{k=1}^{N} S_k(\boldsymbol{r}) \tag{3}$$

Subtracting the reference patterns of all other channels yields the channel-specific residual speckle pattern

$$\tilde{S}_k(\boldsymbol{r})=\left[M(\boldsymbol{r})-\sum_{j\neq k} R_j(\boldsymbol{r})\right]_+ \tag{4}$$

where $[\cdot]_+$ clips negative values to zero. Reference subtraction does not completely eliminate contributions from other channels, because aberrations can alter both speckle positions and intensities. Nevertheless, the resulting residual retains spatial features associated with the target channel, allowing its displacement to be estimated relative to the corresponding reference. We therefore employ a fine-tuned RAFT network [19] to recover the dense displacement field from the reference and residual speckle patterns

$$\boldsymbol{u}_k(\boldsymbol{r})=\left[u_{x,k}(\boldsymbol{r}),u_{y,k}(\boldsymbol{r})\right] \tag{5}$$

Using an all-pairs feature correlation volume, RAFT refines the displacement field in Eq. (5) over five recurrent updates, avoiding the repeated local subset searches required by conventional DIC. Figure 1(b) shows the displacement for a single channel, with the index $k$ omitted from the output label.

Equation (2) relates the speckle displacement to the local wavefront gradient under the small-angle approximation. Conventional numerical integration may propagate local displacement errors and residual interchannel interference into the reconstructed OPD. To bypass numerical integration while enforcing global modal regularity, we employ a Zernike-based regression network that combines displacement-derived features with speckle intensity information to directly estimate the wavefront coefficients. With the pixel arguments omitted, the eight co-registered features are concatenated into the input tensor

$$\boldsymbol{F}_k = \begin{bmatrix} u_{x,k}\ ,\ u_{y,k}\ ,\ |\boldsymbol{u}_k|\ ,\ \nabla_{\mathrm{d}}\cdot\boldsymbol{u}_k\ , \\ \nabla_{\mathrm{d}}\times\boldsymbol{u}_k\ ,\ R_k\ ,\ \widetilde{S}_k\ ,\ |\widetilde{S}_k - R_k| \end{bmatrix} \tag{6}$$

where the entries are the horizontal and vertical pixel displacements, displacement magnitude $|\boldsymbol{u}_k|$, finite-difference divergence $\nabla_{\mathrm{d}}\cdot\boldsymbol{u}_k$, scalar out-of-plane curl $\nabla_{\mathrm{d}}\times\boldsymbol{u}_k$, and intensity channels $R_k$, $\widetilde{S}_k$, $|\widetilde{S}_k - R_k|$. The convolutional network regresses the first $L$ non-piston Zernike coefficients, ordered first by radial order $n$ and then by azimuthal order $m$. For channel $k$, the predicted coefficient vector is

$$\widehat{\boldsymbol{C}}_k = \left[\widehat{C}_{k,1}, \widehat{C}_{k,2}, \ldots, \widehat{C}_{k,L}\right] \tag{7}$$

At normalized pupil coordinates $(\rho,\theta)$, combining the coefficients in Eq. (7) with the Zernike basis functions $Z_l(\rho,\theta)$ gives the reconstructed OPD

$$\widehat{W}_k(\rho,\theta) = \sum_{l=1}^{L} \widehat{C}_{k,l}\, Z_l(\rho,\theta) \tag{8}$$

Here, the modal expansion order $L$ can be flexibly scaled to match the spatial complexity of the target aberrations. In this work, $L = 15$ (up to $Z_5^{-5}$) is adopted.

To map the spatially distributed feature tensor $\boldsymbol{F}_k$ to the compact modal vector $\widehat{\boldsymbol{C}}_k$ while suppressing the influence of regions outside the pupil, the regression network [Fig. 1(c)] employs a multi-stage convolutional encoder with pupil-masked group normalization, sigmoid linear unit (SiLU) activation, and progressive average pooling. This design aggregates multi-scale wavefront features strictly within the active aperture before projecting them through pupil-masked global average pooling (GAP) and a fully connected head (implemented with five encoder blocks of 32–256 channels and a 256–128–$L$ head). Training proceeds in three sequential stages: pre-training the regressor on ground-truth displacements and speckle features, fine-tuning FlyingThings3D-initialized RAFT [22] on simulated displacement labels with recurrently weighted supervision [19], and freezing RAFT to fine-tune the regressor on predicted displacements to accommodate upstream inference errors. The regressor minimizes

$$\mathcal{L} = \alpha\, \mathcal{L}_{\mathrm{coeff}} + \beta\, \mathcal{L}_{\mathrm{OPD}} + \gamma\, \mathcal{L}_{\mathrm{grad}} \tag{9}$$

combining the $L_1$ errors of the Zernike coefficients ($\mathcal{L}_{\mathrm{coeff}}$), pupil-masked OPD ($\mathcal{L}_{\mathrm{OPD}}$), and bidirectional spatial gradients ($\mathcal{L}_{\mathrm{grad}}$) relative to the ground truth. The weighting coefficients can be tuned to balance modal accuracy with fine-scale spatial consistency while preventing the gradient penalty from dominating optimization ($\alpha = 1$, $\beta = 1$, and $\gamma = 0.25$ in our work).

As a proof of concept for the general $N$-channel framework, we simulated $N = 3$ multiplexed channels on a 512 × 512-pixel grid ($\lambda = 525$ nm, $p = 6.5$ μm, and $d = 3$ mm) using 3,000 training, 100 validation, and 100 test samples. To model alignment and illumination variations, we randomly perturbed the nominal equilateral guide-star triangle (1.2° angular radius) in orientation (±15°), radial spacing (±20%), and individual channel angles (±0.1°). Relative channel and overall intensities varied within ±20% and ±10%, respectively, for peak-to-valley (PV) wavefront amplitudes from 0 to 2.5$\lambda$.

Training included noise-free speckle patterns and patterns with Poisson–Gaussian noise. Peak photon counts were sampled logarithmically from 5,000 to 50,000, with Gaussian noise standard deviations up to 0.5% of peak intensity. Each stage used AdamW for 30 epochs with cosine learning-rate annealing. RAFT used six recurrent updates during training and five during inference.

Using channel-specific residuals, we simultaneously reconstructed three wavefronts from one multiplexed speckle measurement (Fig. 2). The OPDs reproduced the main ground-truth features, with a mean channel root-mean-square error (RMSE) of 16.94 nm. Across 100 noise-free test samples, mean OPD RMSE was 19.0 nm, comparable to 20.1 nm for conventional iterative DIC with numerical integration [Fig. 3(a)].

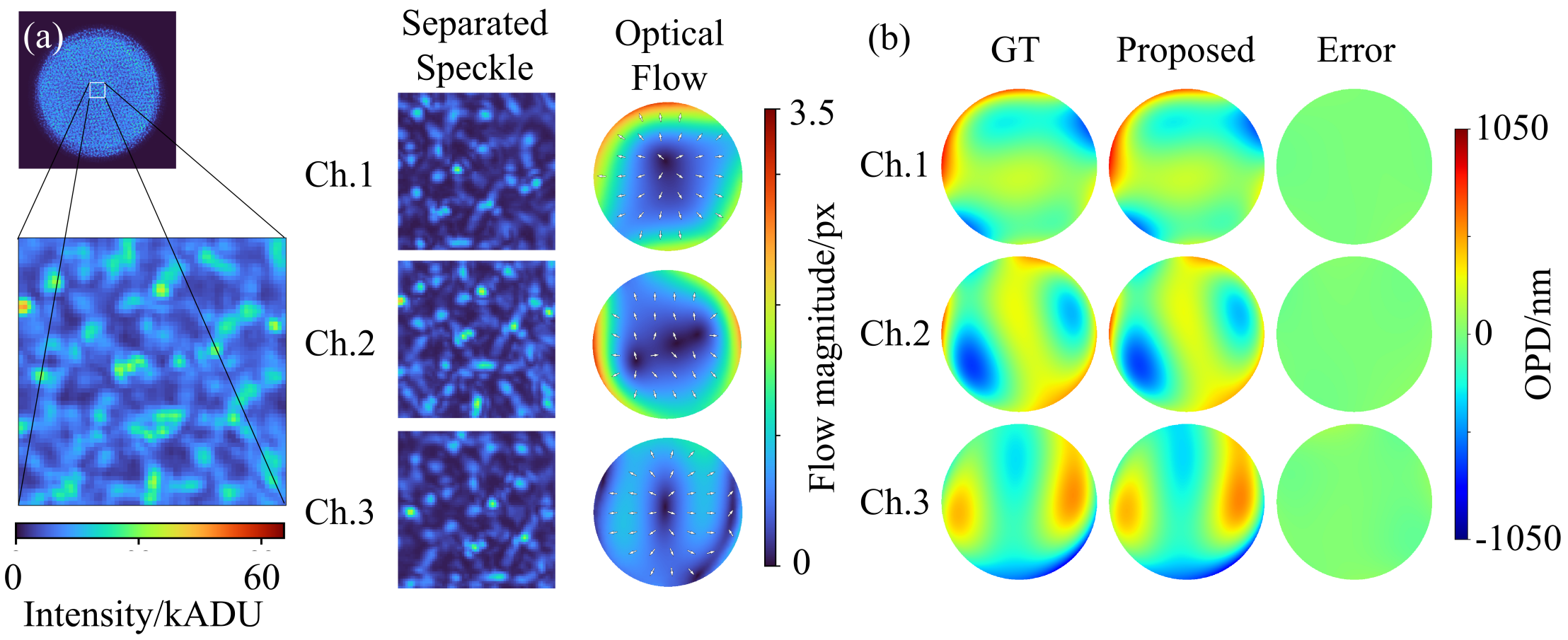

**Fig. 2.** Representative three-channel reconstruction in simulation. (a) Multiplexed speckle image with a 64 × 64-pixel enlargement (white box), channel-specific patterns reconstructed by warping each reference with its RAFT-estimated displacement field, and optical flow for Ch. 1–3 (color indicates magnitude and arrows indicate direction). (b) Ground-truth (GT) and reconstructed OPDs with residual error maps. The OPD maps omit defocus for display only.

Reconstructing three-channel OPDs from preloaded speckle patterns required 23.1 ms on an NVIDIA RTX 4090 GPU, compared with 2801.6 ms for multithreaded MATLAB DIC on an Intel Core i7-14650HX CPU [Fig. 3(b)]. This corresponds to an approximately 121-fold speedup under the tested hardware and software configurations, excluding disk I/O and optical acquisition.

We compared the noise robustness of Zernike regression and numerical gradient integration. Independent zero-mean Gaussian noise was added to the multiplexed and reference images before residual construction, with a standard deviation of $\sigma_n = I_{RMS} 10^{-SNR_{dB}/20}$, where $I_{RMS}$ is the uncentered root-mean-square intensity across all 512 × 512 pixels and $SNR_{dB}$ is the prescribed SNR. Negative intensities were clipped before recomputing residuals and RAFT fields.

Using identical RAFT displacement fields across 100 test samples at SNRs of 10–30 dB, Zernike regression consistently achieved lower OPD RMSE than numerical integration [Fig. 3(c)]. In a representative case with moderate aberration at 15 dB SNR, Zernike regression better preserved the main wavefront features and reduced OPD RMSE from 69.41 to 24.21 nm [Fig. 3(d)].

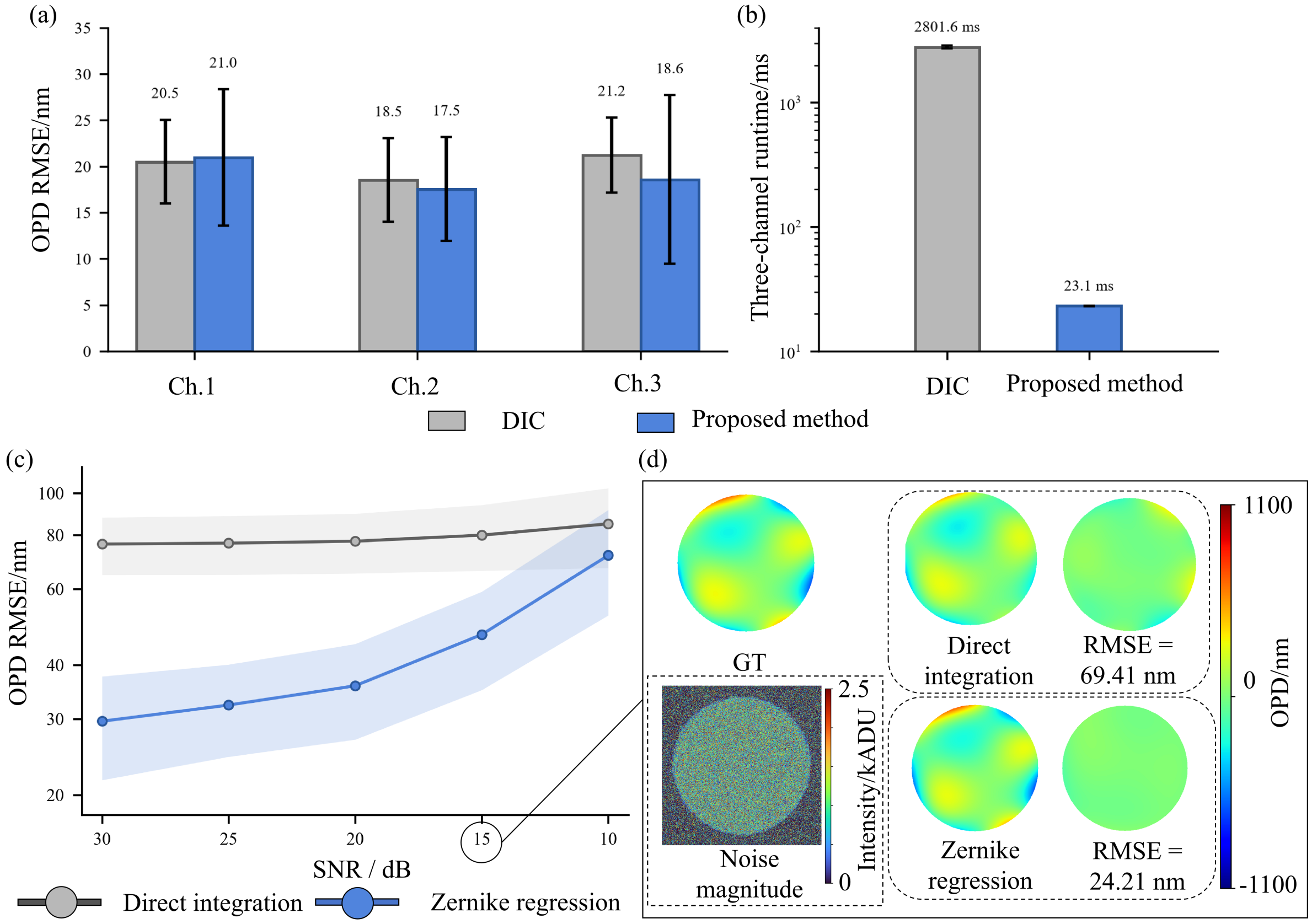

**Fig. 3.** Simulation evaluation of reconstruction accuracy, runtime, and noise robustness. (a) Noise-free channel OPD RMSE (100 test samples). (b) Three-channel runtime (log scale) over ten runs after five warm-ups (error bars: standard deviations). (c) Log-scale OPD RMSE versus SNR (100 test samples; shading indicates standard deviations) comparing Zernike regression and direct integration under identical RAFT fields. (d) Representative reconstructions at SNR = 15 dB, showing GT OPD, noise magnitude, and reconstructed OPDs with residual error maps.

# Results

To evaluate the proposed framework under experimental conditions, we implemented a three-channel wavefront-sensing setup on an Olympus IX-71 inverted microscope. A phase-only spatial light modulator (X13138-01, Hamamatsu) shaped a 532-nm excitation beam into three spots to selectively excite fluorescent-bead guide stars. A 1° holographic diffuser placed near the beads introduced field-dependent aberrations, while a 20×, numerical aperture (NA) = 0.45 objective (LUCPlanFLN, Olympus) collected the fluorescence. The HD-WFS module, conjugated to the objective back pupil, comprised another 1° diffuser (Edmund Optics) placed 3 mm before an sCMOS camera (Zyla 5.5, Andor).

To adapt the networks to experimental speckle characteristics, aberrations in sequentially acquired single-channel speckle patterns were first estimated using conventional DIC and numerically compensated to obtain aberration-corrected reference patterns. Channel-specific references were then generated by translating the corrected reference to the corresponding channel positions. Known Zernike aberrations were subsequently applied to these references to generate synthetic training data incorporating the experimental background, illumination nonuniformity, and acquisition noise characteristics. The resulting synthetic single-channel, multiplexed, and residual patterns provided displacement and coefficient labels for 3,000 training, 100 validation, and 100 test samples, with fixed channel directions and no additional simulated noise. Both networks were fine-tuned using these synthetic data, while experimentally acquired multiplexed measurements were reserved for final evaluation.

In the absence of an independent wavefront ground truth, sequential single-channel DIC reconstructions were used as references. Both multiplexed DIC and the proposed method recovered the main OPD features [Fig. 4(a)]. However, the proposed method showed closer agreement with the sequential references, with channel RMSEs of 17.3, 16.1, and 10.2 nm, compared with 39.2, 17.9, and 24.3 nm for multiplexed DIC [Fig. 4(b)].

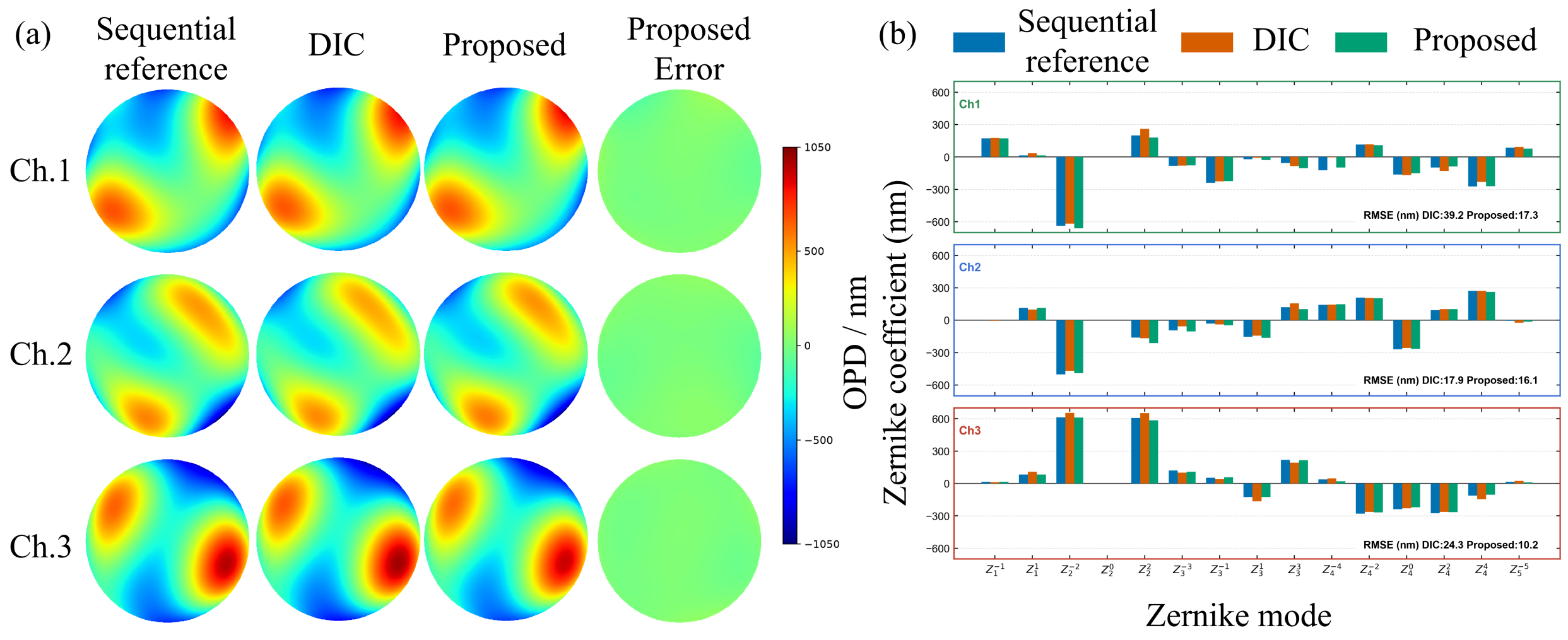

**Fig. 4.** Experimental validation of three-channel multiplexed wavefront sensing. (a) OPD maps from sequential single-channel DIC (reference), multiplexed DIC, and the proposed method. Error maps show the proposed method's deviations from the sequential references. The OPD maps omit defocus for display only. (b) Reconstructed Zernike coefficients and channel RMSEs evaluated against the sequential references.

## Conclusion and Discussion

In conclusion, we propose a deep-learning-based framework for rapid multiplexed wavefront reconstruction from a single speckle pattern, offering a substantial improvement in the measurement efficiency of HD-WFS. By integrating channel-specific residual speckle patterns, RAFT-based displacement field estimation, and Zernike coefficient regression, the proposed framework establishes an efficient and robust reconstruction pipeline that bridges speckle-based optical measurements and data-driven wavefront retrieval. In the three-channel benchmark, reconstruction was approximately 121 times faster than conventional DIC processing. In the noise simulations, Zernike regression yielded lower OPD errors than direct numerical integration when both used the same displacement fields. These results demonstrate the potential of deep learning to overcome the computational bottlenecks of conventional wavefront reconstruction and establish a promising pathway toward real-time, dynamic wide-field aberration sensing and high-speed adaptive optics.

In the current implementation, wavefront reconstruction relies on five iterative RAFT updates, with displacement estimation taking approximately 17.0 ms and accounting for 73.8% of the total reconstruction time. This identifies iterative displacement estimation as the primary computational bottleneck and suggests that further acceleration could be achieved by optimizing the network architecture and iterative update strategy, potentially reducing inference latency without compromising reconstruction accuracy. Related model-based deep learning approaches have also improved temporal resolution in computational microscopy [23].

Regarding scalability, although the proposed framework was experimentally demonstrated using three spatially multiplexed channels, increasing the number of multiplexed channels is not expected to substantially increase reconstruction latency when the channels are processed in parallel as in Ref. [13]. In practice, increasing the number of measured channels reduces speckle contrast, which limits the number of wavefronts that a single speckle pattern can reliably encode [13,24].

Another important direction is to improve reconstruction robustness under low-photon-flux conditions, which is essential for extending the applicability of the proposed framework to photon-limited imaging scenarios. Building on the noise robustness observed in the regression ablation experiments, future efforts could combine phase-mask optimization to enhance energy concentration with noise-aware training strategies that explicitly account for realistic photon statistics and detector noise. Such developments may improve reconstruction reliability under low-SNR conditions, opening up opportunities for wavefront sensing in fluorescence microscopy and faint-target measurements.

**Funding.** Xi'an Science and Technology Association (0959202613053); Shaanxi Science and Technology Department (2024QY-SZX-16); Natural Science Basic Research Program of Shaanxi (2026SYS-SYSZD-104); Key Research and Development Program of Shaanxi Province (2026CY-JJQ-016).

**Disclosures.** The authors declare no conflicts of interest.

**Data availability.** Data underlying the results presented in this paper are not publicly available at this time but may be obtained from the authors upon reasonable request.

# References

1. K. M. Hampson, R. Turcotte, D. T. Miller, *et al.*, “Adaptive optics for high-resolution imaging,” Nat. Rev. Methods Primers **1**, 68 (2021).
2. M. J. Booth, “Wave front sensor-less adaptive optics: a model-based approach using sphere packings,” Opt. Express **14**, 1339–1352 (2006).
3. M. Žurauskas, I. M. Dobbie, R. M. Parton, *et al.*, “IsoSense: frequency enhanced sensorless adaptive optics through structured illumination,” Optica **6**, 370–379 (2019).
4. B. C. Platt and R. Shack, “History and principles of Shack-Hartmann wavefront sensing,” J. Refract. Surg. **17**, S573–S577 (2001).
5. K. Wang, W. Sun, C. T. Richie, *et al.*, “Direct wavefront sensing for high-resolution in vivo imaging in scattering tissue,” Nat. Commun. **6**, 7276 (2015).
6. J. Mertz, H. Paudel, and T. G. Bifano, “Field of view advantage of conjugate adaptive optics in microscopy applications,” Appl. Opt. **54**, 3498–3506 (2015).
7. R. D. Simmonds and M. J. Booth, “Modelling of multi-conjugate adaptive optics for spatially variant aberrations in microscopy,” J. Opt. **15**, 094010 (2013).
8. E. Marchetti, R. Brast, B. Delabre, *et al.*, “On-sky testing of the multi-conjugate adaptive optics demonstrator,” The Messenger **129**, 8–13 (2007).
9. D. Ancora, T. Furieri, S. Bonora, *et al.*, “Spinning pupil aberration measurement for anisoplanatic deconvolution,” Opt. Lett. **46**, 2884–2887 (2021).
10. T. Furieri, D. Ancora, G. Calisesi, *et al.*, “Aberration measurement and correction on a large field of view in fluorescence microscopy,” Biomed. Opt. Express **13**, 262–273 (2022).
11. X. Wei and L. Thibos, “Design and validation of a scanning Shack Hartmann aberrometer for measurements of the eye over a wide field of view,” Opt. Express **18**, 1134–1143 (2010).
12. P. Berto, H. Rigneault, and M. Guillon, “Wavefront sensing with a thin diffuser,” Opt. Lett. **42**, 5117–5120 (2017).

13. T. Wu, M. Guillon, G. Tessier, *et al.*, “Multiplexed wavefront sensing with a thin diffuser,” Optica **11**, 297–304 (2024).
14. B. Pan, K. Qian, H. Xie, *et al.*, “Two-dimensional digital image correlation for in-plane displacement and strain measurement: a review,” Meas. Sci. Technol. **20**, 062001 (2009).
15. S. Boukhtache, K. Abdelouahab, F. Berry, *et al.*, “When deep learning meets digital image correlation,” Opt. Lasers Eng. **136**, 106308 (2021).
16. X. Duan, H. Xu, R. Dong, *et al.*, “Digital image correlation based on convolutional neural networks,” Opt. Lasers Eng. **160**, 107234 (2023).
17. Z. Qiao, X. Shi, Y. Yao, *et al.*, “Real-time X-ray phase-contrast imaging using SPINNet—a speckle-based phase-contrast imaging neural network,” Optica **9**, 391–398 (2022).
18. S.-K. Zhu, Z.-H. Zheng, W. Meng, *et al.*, “Harnessing disordered photonics via multi-task learning towards intelligent four-dimensional light field sensors,” PhotoniX **4**, 26 (2023).
19. Z. Teed and J. Deng, “RAFT: Recurrent All-Pairs Field Transforms for Optical Flow,” in Proc. Eur. Conf. Comput. Vis. (ECCV), 402–419 (2020).
20. K. Wang, L. Song, C. Wang, *et al.*, “On the use of deep learning for phase recovery,” Light Sci. Appl. **13**, 4 (2024).
21. Y. Nishizaki, M. Valdivia, R. Horisaki, *et al.*, “Deep learning wavefront sensing,” Opt. Express **27**, 240–251 (2019).
22. N. Mayer, E. Ilg, P. Häusser, *et al.*, “A Large Dataset to Train Convolutional Networks for Disparity, Optical Flow, and Scene Flow Estimation,” in Proc. IEEE Conf. Comput. Vis. Pattern Recognit. (CVPR), 4040–4048 (2016).
23. Y. Gao and L. Cao, “Model-based deep learning enables time-resolved computational microscopy,” PhotoniX **7**, 3 (2026).
24. B. Blochet, N. Lebas, P. Berto, *et al.*, “Single-shot hyperspectral wavefront imaging,” Nat. Commun. **17**, 137 (2026).